\documentclass[twocolumn,preprintnumbers,amsmath,amssymb,prl,showpacs]{revtex4}

\usepackage[]{graphicx}
\usepackage{color}

\usepackage{dcolumn}
\usepackage{bm}

\begin{document}

\title{Caging dynamics in a granular fluid}

\author {P.M. Reis\footnote{Current address: Laboratoire PMMH (UMR7636 CNRS-ESPCI-P6-P7), 10 rue Vauquelin, 75231 Paris, FRANCE. \texttt{preis@pmmh.espci.fr} }, R.A. Ingale and M.D. Shattuck}
\affiliation{ Benjamin Levich Institute, The City College of the
City University of New York \\ 140th St. and Convent Av., New York
NY 10031, USA }

\begin{abstract}
We report an experimental investigation of the caging motion in a
uniformly heated granular fluid, for a wide range of filling fractions,
$\phi$. At low $\phi$ the classic diffusive behavior of a fluid is
observed. However, as $\phi$ is increased, temporary cages develop and
particles become increasingly trapped by their neighbors. We
statistically analyze particle trajectories and observe a number of
robust features typically associated with dense molecular liquids and colloids. Even though our monodisperse and
quasi-2D system is known to not exhibit a glass transition, we still
observe many of the precursors usually associated with glassy
dynamics. We speculate that this is due to a process of structural
arrest provided, in our case, by the presence of crystallization.
\end{abstract}

\pacs{find pacs}

\maketitle
 
In additional to being of great industrial and geological importance, granular materials are of fundamental interest due to their strong non-equilibrium nature \cite{granular:general}. Ensembles of granular particles are intrinsically dissipative and any dynamical study must involve energy injection. These ingredients make the understanding of granular assemblies a challenging endeavor and a general theoretical framework is still lacking. A possible approach is the study of non-equilibrium steady states, in  systems where energy injection perfectly balances dissipation. Whereas some progress has been made in the fast dilute regime \cite{goldhirsch:2003}, the understanding of the dense case of these steady states f remains an open question. One avenue of research has been to borrow concepts from other dense assemblies of particles such as colloids, suspensions and emulsions. Moreover, it is thought that connections with molecular glass formers may be relevant \cite{liu:1998}. \emph{A priori}, it is far from obvious that such analogy may be valid due to the enormous differences in  lengthscales and the the mechanism of energy supply. However, one common feature in all of these seemingly disparate systems is the presence of \emph{cages}: each particle is temporarily trapped by its neighbors and then moves in short bursts due to nearby cooperative motion. This results in highly heterogeneous motion and slowing down of the dynamics. In molecular systems this behavior is typically observed indirectly from scattering experiments \cite{light:scattering}. In colloids, however, caging motion has been observed directly through microscopy, in both 3D \cite{weeks} and quasi-2D \cite{konig:2005} geometries. A large number of theoretical \cite{gotze:1989,cugliandolo:2003} and numerical \cite{barrat:1990,hurley:2005:Zangi:2004} studies have set out to further investigate the importance of this heterogeneous dynamics. The relevance of the \emph{caging} in driven granular materials \cite{pouliquen:2003,dauchot:2005} and air-fluidized particle systems \cite{abate:2006} has only recently started to be addressed. In particular, Dauchot et. al. \cite{dauchot:2005} have reported on a granular system driven by cyclic shear where they observed many ``glasslike'' features, but for a single value of the filling fraction.
 
In this Letter we present a novel dynamical study of a monodisperse, quasi-2D granular fluid \cite{reis:crystallization:2006}, in which the particles are
excited by a spacially uniform stocastic forcing. In our system the
filling fraction, $\phi$, can be varied from a single particle to
hexagonal close packing. We have shown that the structural
configurations of this non-equilibrium steady state are the same as
those of equilibrium hard-disks \cite{reis:crystallization:2006}. In
particular, the system exhibits three phases: an isotropic fluid phase $\phi <
\phi_{l}=0.652$, a crystaline solid phase $\phi > \phi_{s}=0.719$, and
an intermediate phase $\phi_{l}<\phi<\phi_{s}$ consistent with a
hexatic phase\cite{jaster:2004}.  The phase boundaries, $\phi_{l}$ (\emph{liquidus} point) and $\phi_{s}$ (\emph{solidus point}) are determined by structure alone \cite{reis:crystallization:2006}. Here, we explore the dynamics of these phases through single particle trajectories, focusing on the caging dynamics seen in the intermediate phase.
     \begin{figure}[b]
          \begin{center}
    \includegraphics[width=\columnwidth]{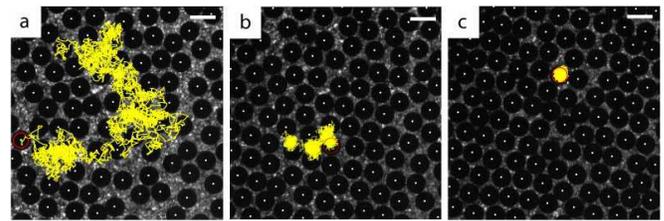}
          \caption{Experimental frames with  superposed typical trajectories of a single particle: (a) $\phi=0.567$, (b) $\phi=0.701$ and (c) $\phi=0.749$. Note that even though only a single trajectory is shown for each $\phi$, particle tracking and statistics were performed over all particles within the imaging window. The scalebar is 2mm.\label{fig:trajectories}} 
          \end{center}          
     \end{figure}
In Fig. \ref{fig:trajectories} we present typical single particle
trajectories for filling fractions in each of the three phases. Simple
fluid behavior is observed at low $\phi$, characterized by random
diffusion (Fig. \ref{fig:trajectories}a). Above crystallization
($\phi>\phi_{s}$) particles become fully arrested by thier six
hexagonally packed neighbors (Fig. \ref{fig:trajectories}c). In the
intermediate phase, we see a mixture of these behaviors
(Fig. \ref{fig:trajectories}b).  At short times, particles are
temporarily trapped in cages formed by thier neighbors, but at long
times they diffuse from cage to cage. We will use 
the Mean Square Displacement (MSD) and the Intermediate Scattering Function (ISF)
to show that the caging dynamics seen here is
qualitatively identical to that of dense molecular and colloidal systems
\cite{weeks,light:scattering} and supercooled liquids \cite{supercooled}. This is surprising since our experiment is fundamentally different. The associated lengthscales of our granular system are typically 3-7 orders of magnitude larger than those of molecular systems and colloids. Moreover, the steady states we study are inherently far from equilibrium since energy is both injected (through vibration) and dissipated in inter-particle inelastic collisions and frictional contacts.

Our experimental apparatus consists of vertically vibrated, $D=1.19mm$
diameter stainless steel spheres confined between two $85.3D$ diameter
horizontal glass plates, separtated by $1.6D$, which constrain the
particle motion to be quasi-2D. Our system is described in more detail in
\cite{reis:crystallization:2006,reis:velocities:2006} and improves on
Olafsen and Urbach's similar system \cite{olafsen:1998} by using a
roughned bottom plate.  This allows us to study a wide range of filling fractions 
($1.4 \times 10^{-4}<\phi<0.8$).  We sinusoidally vibrate the system with a
frequency $f=50Hz$ and a maximum acceleration equal to 4 times
gravity, but the structure and dynamics of the system are independent
of the forcing for a wide range of parameters
\cite{reis:velocities:2006}. To ensure repeatable initial conditions,
we start with all particles hexagonally packed near the boundary.  A
steady-state is reached by waiting for 12000 vibration cycles before
10 seconds of data are acquired.  We record the dynamics in
a ($15\times15mm^{2}$) central region using a high-speed camera at 840
Hz and track the trajectories of all particles.

    \begin{figure}[t]
          \begin{center}
    \includegraphics[width=0.8\columnwidth]{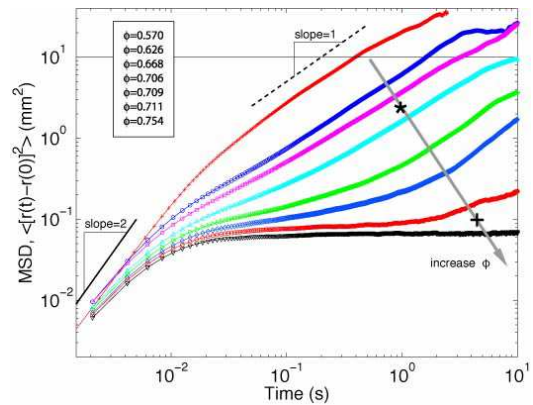}
          \caption{Time dependence of the Mean Square Displacement for various values  of filling fraction (numerical values given in the box). The curve marked with a square is for a single particle in the cell. The arrow points in the direction of increasing $\phi$. Along the arrow, the symbols (*) and (+) are located at $\phi_{l}$ and $\phi_{s}$, respectively. The horizontal line corresponds to the area of 1/4th of the system, above which finite system size effects become important. \label{fig:msd}}
          \end{center}          
     \end{figure}

We first analyse the particle trajectories by measuring the Mean Square Displacement defined as, 
\begin{equation}
	M(t)=\langle [r(t) - r(0)]^{2}\rangle,
\end{equation}
where $r(t)$ is the position of a particle at time $t$, $r(0)$ is
its initial position, the brackets $\langle \rangle$
signify ensemble averaging over many realizations and time invariance
is assumed. The MSD for a range of $\phi$ are shown in
Fig. \ref{fig:msd}. For the case of a single particle in the cell
(marked with a square in Fig. \ref{fig:msd}), the motion at short times is ballistic, and
$M(t)\sim t^{\alpha}$ where $\alpha\sim2$ ($\alpha=2$ for pure
ballistic motion). At later times, the particle moves diffusively and
the slope of $M(t)$ tends to $\alpha=1$. This shows that the
trajectory of a single particle is indeed randomized across the
cell. For all $\phi$, the motion at early times is superdiffusive with
$\alpha\sim2$ showing ballistic motion. For $\phi<0.719$ the motion
always becomes diffusive at long times with $M(t)\sim t$. Eventually, this becomes increasingly noisy due to the lack of statistics in the time averaging. For
$\phi>0.719$ the particles are trapped by their six hexagonally packed
neighbors, and $M(t)$ levels off to a constant value set by the
lattice spacing. In the intermediate phase, however, a plateau emerges
at intermediate times where the motion is subdiffusive with
$0<\alpha<1$. This plateau becomes obvious above the liquidus point,
$\phi_{l}$ (marked as * in Fig. \ref{fig:msd}) and represents the
slowing down due to the cage effect shown in
Fig. \ref{fig:trajectories}b. A similar dependence of $M(t)$ has recently been observed in a quasi-2D system of bidispersed particles fluidized by a uniform upflow or air \cite{abate:2006}.

Another classic measure in the study of dense liquid phases is the
Intermediate Scattering Function  \cite{light:scattering} which is
defined as,
\begin{equation}
		F_{s}(\mathbf{q},t)=\frac{1}{N}\sum_{j}
		\langle
		\exp \left(
			-i \mathbf{q}\cdot [ \mathbf{r}_{j}(t) - \mathbf{r}_{j}(0) ]		\right)
		\rangle,
	\label{eqn:isf}
\end{equation}
where $\mathbf{q}$ is a wavenumber and $\mathbf{r}_{j}(t)$ is the
trajectory of particle $j$ out of $N$ particles in the system. This
measure is widely used in colloids since it is readily available
through light scattering experiments \cite{light:scattering} and is,
essentially, a measure of the time decorrelation of the positional
wavevectors. In dense colloids and supercooled liquids, 
$F_{s}(t,q)$ captures the relaxation due to caging in the form of a
two-step relaxation: 1) the fast (early time) $\beta$ relaxation which
corresponds to the diffusion inside the cage followed by 2) the
$\alpha$ relaxation corresponding to the time it takes for the
particle to diffuse out of the cage.

     \begin{figure}[t]
          \begin{center}
    \includegraphics[width=0.8\columnwidth]{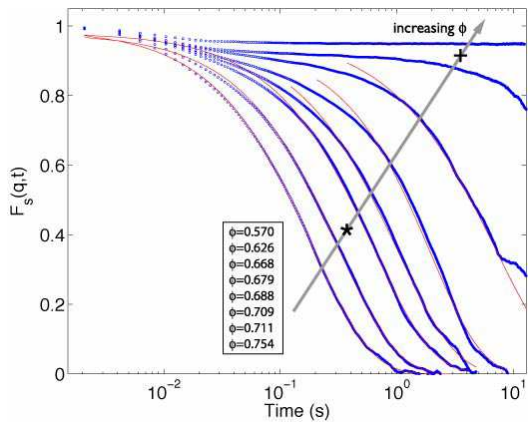}
          \caption{Time dependence of the Intermediate Scattering
          Function, with $qD=2.14$, for various filling fractions (numerical values of
          $\phi$ given in the box). The arrow points in the direction
          of increasing $\phi$. Along the arrow, the symbols (*) and
          (+) are located at $\phi_{l}$ and $\phi_{s}$,
          respectively. The solid lines are fits to
          Eqn. (\ref{eqn:stretchedexponential}). 
          \label{fig:isf_phi}}
          \end{center}          
     \end{figure}

     \begin{figure}[b]
          \begin{center}
          \includegraphics[width=0.7\columnwidth]{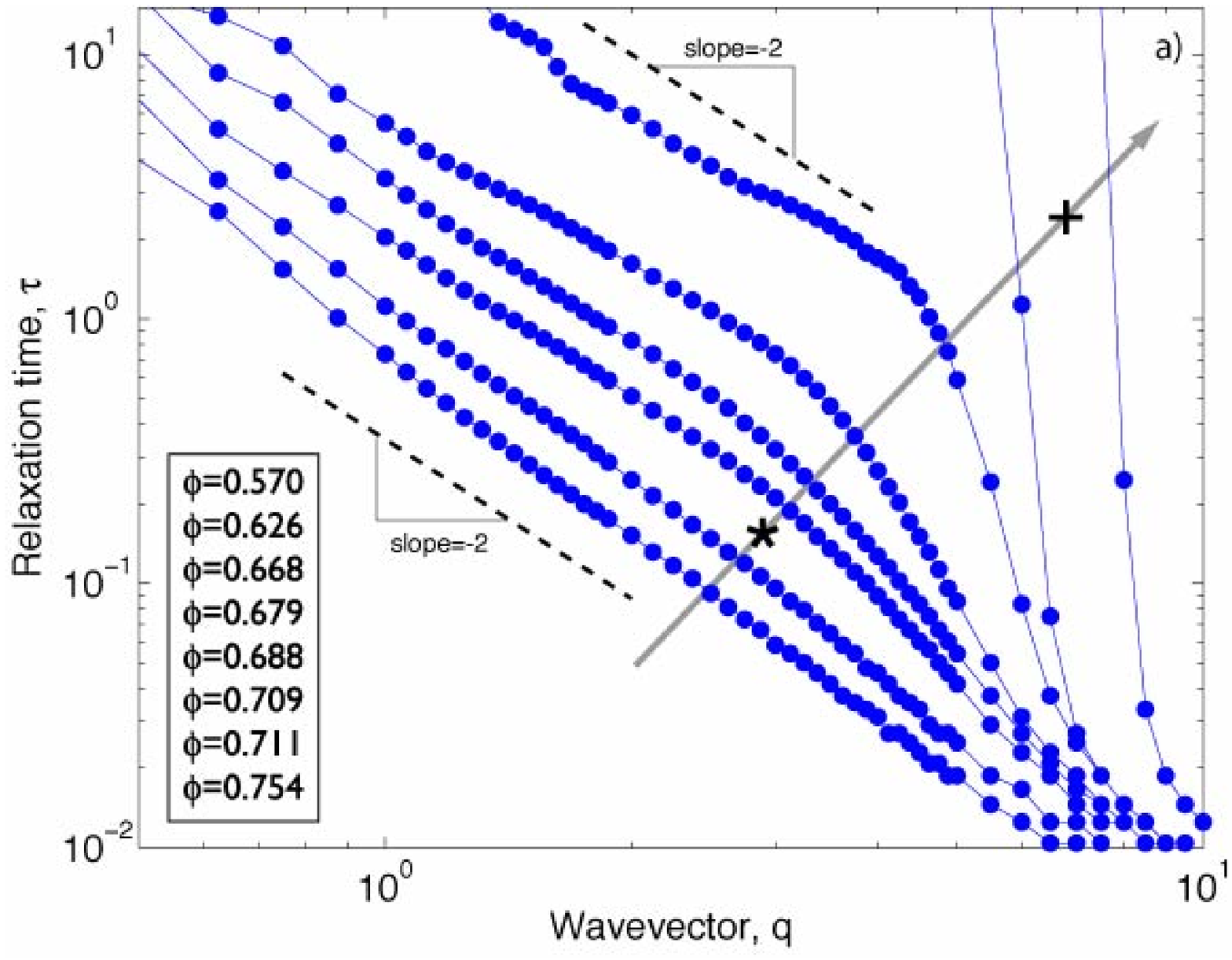}
          \includegraphics[width=0.7\columnwidth]{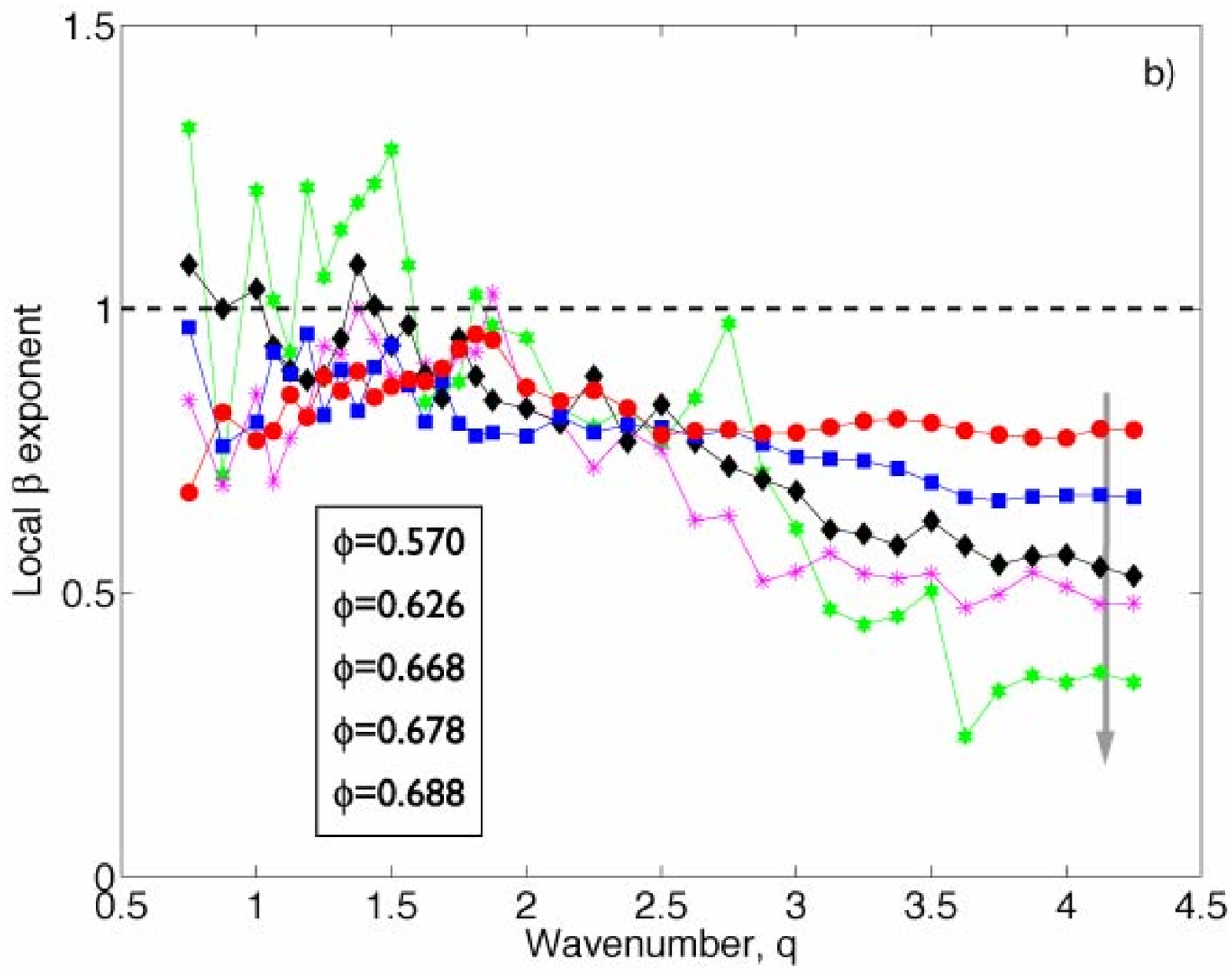}
          \caption{(a) Wavevector dependence of the relaxation time, $\tau$ and (b) local stretching exponent $\beta$, for various values of filling fraction. The arrows point in the direction of increasing $\phi$ and the numerical values of $\phi$ are given in the boxes. Along the arrow, the symbols (*) and (+) are located at $\phi_{l}$ and $\phi_{s}$, respectively.           
          \label{fig:tau_beta}}
          \end{center}          
     \end{figure}    

In Fig. \ref{fig:isf_phi} we plot $F_{s}(t)$ for wavevector
$qD=2.14$ ($q=1.8mm^{-1}$) for various values of $\phi$. As expected,
in the crystal phase ($\phi>0.719$) $F_{s}$ levels at a value close to
1 and little decorrelation occurs, since each particle is fully
trapped. In the fluid phase, $F_{s}$ rapidly decays as particles move
across the cell diffusively and the initial positional wavevector
quickly decorrelates. In the intermediate phase we observe the classic
$\alpha$ and $\beta$ two-step relaxation; there is  a clear intermediate
plateau and the $\alpha$ relaxation occurs at increasingly longer
timescales, as $\phi$ is increased. As for the MSD before, this
two-step relaxation becomes visible above the liquidus point,
$\phi_{l}$ (marked as * in Fig. \ref{fig:isf_phi}). For each value of $\phi$, the $\alpha$ relaxation at late times is well described by a stretched exponential of the form,
	\begin{equation}
		F_{s}(q,t)\sim \exp [ - \left( t /\tau(q)\right)^{\beta(q)} ],
	\label{eqn:stretchedexponential}	
	\end{equation}
where $\tau(q)$ is a relaxation time  and the stretching exponent is
typically $\beta(q)\le1$. Fits of this stretched exponential form to
the experimental data are shown as solid lines in
Fig. \ref{fig:isf_phi}. It is interesting to note that this behavior is in good agreement with the predictions of Mode Coupling Theory (MCT) \cite{gotze:1989}. This is highly surprising since MCT has been developed for thermal fluids and is not known to apply to non-equilibrium systems such as ours.

We now focus on the $q$-dependence of both  the relaxation time $\tau(q)$ (Fig. \ref{fig:tau_beta}a)  and the exponent $\beta(q)$ (Fig. \ref{fig:tau_beta}b), for various values of $\phi$. We defined $\tau(q)$ as the time it takes for the experimental curves of $F_{s}$ to fall to a level of $1/e$, i.e. $F_{s}(\tau)=1/e$. The stretching exponent $\beta(q)$ is the local slope of the quantity $\log (-\log(F_{s}))$, in the neighborhood of $\tau$.  At low filling fractions, $\tau$ scales as $q^{{-2}}$. As the filling fraction is increased this scaling
continues to hold but only up to a cut off value above which
(lengthscale below which) it sharply drops. This cutoff lengthscale
can be associated with a characteristic size of the cage, which
becomes increasingly smaller as $\phi$ is increased. On the other hand
$\beta$, the local exponent, tends to one at small $q$ but decreases
progressively below one for higher filling fractions and $F_{s}$
becomes increasingly stretched. These findings can now be combined
and interpreted as follows. For small $q$ (i.e. large lengthscales),
$\tau\sim q^{-2}$ and $\beta(q)\rightarrow 1$, which together imply
that $F_{s}(q,t)\sim \exp (-Dq^{2}t)$. This is the result for
classical diffusion \cite{dauchot:2005}. For large $q$ (i.e. small
lengthscales) this Brownian scaling breaks down to a stretched
exponential with $\beta <1$, which can attributed to the presence of
dynamic heterogeneities due to caging.

     \begin{figure}[t]
          \begin{center}
    \includegraphics[width=\columnwidth]{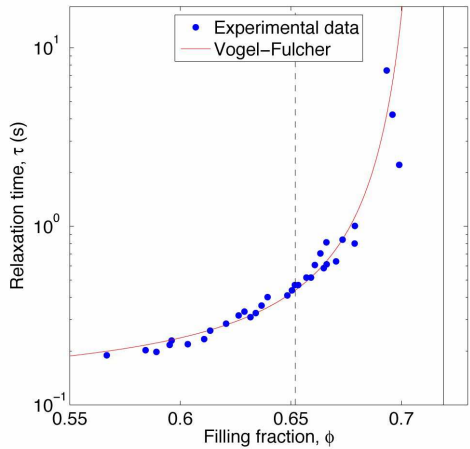}
          \caption{Relaxation time, $\tau$, extracted from
          the Intermediate Scattering Function as a function of filling
          fraction. The solid line is a fit to the Vogel-Fulcher law
          of Eqn. (\ref{eqn:vogel:fulcher}). Dashed and solid lines
          represent the location of the liquidus and solidus points,
          respectively.\label{fig:tau_phi}}
          \end{center}          
     \end{figure}

Returning to the case of fixed $qD=2.14$, we plot $\tau$ as a function
of $\phi$ in Fig. \ref{fig:tau_phi}. A significant slowing down
of the dynamics can be seen at high $\phi$, as crystallization is
approached. It is highly surprising that this slowing down with $\phi$ is well described by the Vogel-Fulcher law \cite{vogel:fulcher} as found in many glass forming systems,
\begin{equation}
\tau\sim \exp [ A/(\phi_{c} - \phi) ],
\label{eqn:vogel:fulcher}
\end{equation}
where $A=0.094\pm0.004$ is a fitting parameter but
$\phi_{c}=0.719\pm0.007$ is the filling fraction for crystallization
which was determined independently from experiments
\cite{reis:crystallization:2006}. A power-law fit was not as
satisfactory. Note that, as the fluid goes through the transition from isotropic fluid  to the
intermediate phase (at $\phi_{l}=0.652$), $\tau$ shows no
particular feature. This functional dependence of the relaxation has
also recently been found within the granular context by Fierro
\emph{et. al.} \cite{fierro:2005} in a numerical lattice model.

In summary, we have studied the dynamics of a uniformly heated
granular fluid. We have observed a number of features typically
associated with dense liquid behavior in molecular and colloidal systems, namely: prominence of cages, development of a plateau in the MSD and ISF with the breakdown of the Brownian diffusive behavior and a Vogel-Fulcher relaxation. These features become particularly visible above the liquidus
point, for $\phi\geq\phi_{l}$. In particular, our results can be directly compared to both experiments \cite{konig:2005} and simulations \cite{hurley:2005:Zangi:2004} of quasi-2D colloidal particles. This is surprising since our experiment is fundamentally different: it is intrinsically far from equilibrium since energy is not conserved, the constituent particles are macroscopic  and it is known that for  monodisperse and quasi-2D systems such as ours there is no ideal glass transition \cite{krauth:2000:donev:2006} (the phase diagram is fluid, intermediate phase and crystal).  In the absence of a structural glass, we propose that the spatially uniform stochastic way of injecting energy along with a process of structural arrest provided by crystallization \cite{reis:crystallization:2006} are at play to account for the many observed similarities. This suggests that theoretical frameworks previously developed for dense thermal liquids, for example MCT \cite{gotze:1989}, might shed some light to the description of excited granular materials.

This work is funded by The National Science Foundation, Math, Physical
Sciences Department of Materials Research under the Faculty Early
Career Development (CAREER) Program (DMR-0134837). PMR was partially
funded by the Portuguese Ministry of Science and Technology under the
POCTI program and the MECHPLANT NEST-Adventure program of the European
Union.


\begin{thebibliography}{99}

\bibitem{granular:general} H. Jaeger and S. Nagel \emph{Science} \textbf{255}, 1523 (1992). H. Jaeger, S. Nagel and R. Behringer \emph{Rev. Mod. Phys.} \textbf{68}, 1259(1996).

\bibitem{goldhirsch:2003} I. Goldhirsch, \emph{Ann. Rev. of Fluid Mech.} \textbf{267} 267 (2003).  


\bibitem{liu:1998} A.J. Liu and S.R. Nagel, \emph{Nature} \textbf{396} 21 (1998).  

\bibitem{light:scattering} P.N. Pusey and W. van Megen \emph{Physica A} \textbf{157}, 705 (1989).  W. van Megen and S.M. Underwood \emph{Phys. Rev. E} \textbf{47}, 248 (1993).

\bibitem{weeks} E.R. Weeks, J.C. Crocker, A.C. Levitt, A. Schofield and D.A. Weitz, \emph{Science} \textbf{287}, 627 (2000).

\bibitem{konig:2005} H. K\"{onig}, R. Hund, Z. Zahn and G. Maret \emph{Eur. Phys. J. E} \textbf{18}, 287 (2005). 


\bibitem{gotze:1989} W. G\"otze, in \emph{Liquids, Freezing and Glass Transition}, edited by J. P. Hansen, D. Levesque, and J. Zinn-Justin (North Holland, Amsterdam, 1991), Les Houches Summer Schools of Theoretical Physics Session LI 287 (1989).

\bibitem{cugliandolo:2003} L. F. Cugliandolo, in \emph{Slow Relaxations and nonequilibrium dynamics in condensed matter}, edited by J.-L. Barrat, M. Feigelman, J. Kurchan and J. Dalibard (Springer Berlin/Heidelberg) Les Houches Summer School,  \textbf{77} 367 (2004).

\bibitem{barrat:1990} J.-L. Barrat, J.-N. Roux and J.-P. Hansen \emph{Chem. Phys.} \textbf{149}, 197 (1990).

\bibitem{hurley:2005:Zangi:2004} M. M. Hurley and P. Harrowell \emph{Phys. Rev. E} \textbf{52}, 1694 (1995). R. Zangi and S.A. Rice \emph{Phys. Rev. Lett.} \textbf{92}, 035502 (2004).
 
\bibitem{pouliquen:2003} O. Pouliquen, M. Belzons and M. Nicolas, \emph{Phys. Rev. Lett.} \textbf{91}, 014301 (2003).

\bibitem{dauchot:2005} G. Marty and O. Dauchot, \emph{Phys. Rev. Lett.} \textbf{94}, 015701 (2005). O. Dauchot, G. Marty and G. Biroli, \emph{Phys. Rev. Lett.}  \textbf{95}, 265701 (2006).

\bibitem{abate:2006} A.R. Abate and D.J. Durian, \emph{Phys. Rev. E}, \textbf {74}, 031308 (2006).

\bibitem{reis:crystallization:2006} P.M. Reis, R.A. Ingale and M.D. Shattuck, \emph{Phys. Rev. Lett.} \textbf{96}, 258001 (2006). 

\bibitem{supercooled} H. Sillescu,  \emph{J. Non-Crystal. Solids} \textbf{243} 81 (1999). M. D. Ediger, \emph{Annu. Rev. Phys. Chem.} \textbf{51}, 99 (2000).

\bibitem{reis:velocities:2006} P.M. Reis, R.A. Ingale and M.D. Shattuck, in preparation for \emph{Phys. Rev. E} (2006).

\bibitem{jaster:2004} A. Jaster, \emph{Phys. Lett. A} \textbf{330}, 120 (2004).

\bibitem{krauth:2000:donev:2006}  L. Santen and W. Krauth, \emph{Nature} \textbf{405}, 550 (2000). A. Donev, and F.H. Stillinger, and S. Torquato \emph{Phys. Rev. Lett.} \textbf{96}, 225502 (2006). 

Nature 405, 550-551(1 June 2000)

\bibitem{olafsen:1998} J.S. Olafsen and J.S. Urbach, \emph{Phys. Rev. Lett.}, \textbf{81}, 4369 (1998).



\bibitem{vogel:fulcher} H. Vogel \emph{Z. Phys.} \textbf{22}, 645 (1921). G.S. Fulcher \emph{J. Am. Ceram. Soc.} \textbf{6}, 339 (1925).



\bibitem{fierro:2005} A. Fierro, M. Nicodemi, M. Tarzia, A. de Candia and A. Coniglio \emph{Phys. Rev. E} \textbf{71}, 061305 (2005).

\end{thebibliography}
\end{document}